\documentstyle[12pt,epsf,epsfig,wrapfig]{article}
\begin{document}
\begin{center}
{\bfseries AZIMUTHAL ASYMMETRIES IN INCLUSIVE SINGLE 
PION ELECTROPRODUCTION~\footnote{Invited talk at Spin-01, August 2 -- 7, JINR Dubna, 
Russia.}}
\vskip 5mm   
K.A. Oganessyan$^{\dag}$
\vskip 5mm
{\small
{\it
INFN-Laboratori Nazionali di Frascati I-00044 Frascati, 
via Enrico Fermi 40, Italy
}
\\
{\it
DESY, Deutsches Elektronen Synchrotron 
Notkestrasse 85, 22603 Hamburg, Germany              
}
\\
$\dag$ {\it
E-mail: kogan@mail.desy.de
}}
\end{center}
\vskip 5mm     
\begin{abstract}
The leading and sub-leading order results 
for pion electroproduction in polarized and unpolarized semi-inclusive 
deep-inelastic scattering, are considered putting emphasis on transverse 
momentum dependent effects appearing in azimuthal asymmetries. 
In particular the spin-dependent (single, double) and spin-independent 
asymmetries of the distributions in the azimuthal angle $\phi$ 
of the pion related to the lepton scattering plane are discussed. 
\end{abstract}

\vskip 10mm

\section{Introduction}

Semi-inclusive deep inelastic scattering (SIDIS) of leptons off a nucleon 
is an important process to study the internal structure of the 
nucleon and its spin properties. In particular, measurements of 
azimuthal distributions of the detected hadron provide valuable 
information on hadron structure functions, quark-gluon correlations 
and parton fragmentation functions. 

The kinematics of SIDIS in the case of longitudinally 
polarized target is illustrated in Fig.1: $k_1$ ($k_2$) is the 
4-momentum of the 
incoming (outgoing) charged lepton, $Q^2=-q^2$, where $q=k_1-k_2$ 
is the 4-momentum of the virtual photon. 
$P$ ($P_h$) is the momentum of the target (observed) hadron, $x=Q^2/2(Pq)$, 
$y=(Pq)/(Pk_1)$, $z=(PP_h)/(Pq)$, $k_{1T}$ is the incoming lepton
transverse momentum with respect to the virtual photon momentum direction, 
and $\phi$ is the azimuthal angle between $P_{hT}$ and $k_{1T}$ around the 
virtual photon direction, angle $\theta_{\gamma}$ is a virtual photon 
emission angle and $S_{lab}$ is the target polarization parallel to the 
incoming lepton momentum, $S_L$ and $S_{Tx}$ are  the longitudinal and transverse 
spin in the virtual photon frame, respectively~\cite{OABK}:     
\begin{equation}
\label{ST0}
S_L = S_{lab} \cos{\theta}_{\gamma}, \quad S_{Tx} = S_{lab} \sin{\theta}_{\gamma},  
\end{equation}
\begin{equation}
\label{ST}
\sin{\theta}_{\gamma} = \sqrt{\frac{4M^2\,x^2}{Q^2+4M^2\,x^2}
    (1-y-{M^2x^2\,y^2 \over Q^2})}.
\end{equation}

In SIDIS one assumes the factorization of the cross section, schematically
\begin{equation}
d\sigma^{lN \to l'hX}=\sum _q f^{H \to q}\otimes d\sigma^{eq \to eq} \otimes 
D^{q \to h}, 
\label{R2}
\end{equation} 
where the soft parts, the distribution function $f$ and the fragmentation 
function $D$ depend not only on $x$ and $z$, respectively, but also on 
quark's transverse momenta; $d \sigma^{eq \to eq}$ describes the scattering 
among elementary constituents and can be calculated perturbatively in the 
framework of quantum chromodynamics (QCD). 

\begin{wrapfigure}{l}{8.0cm}
\epsfig{figure=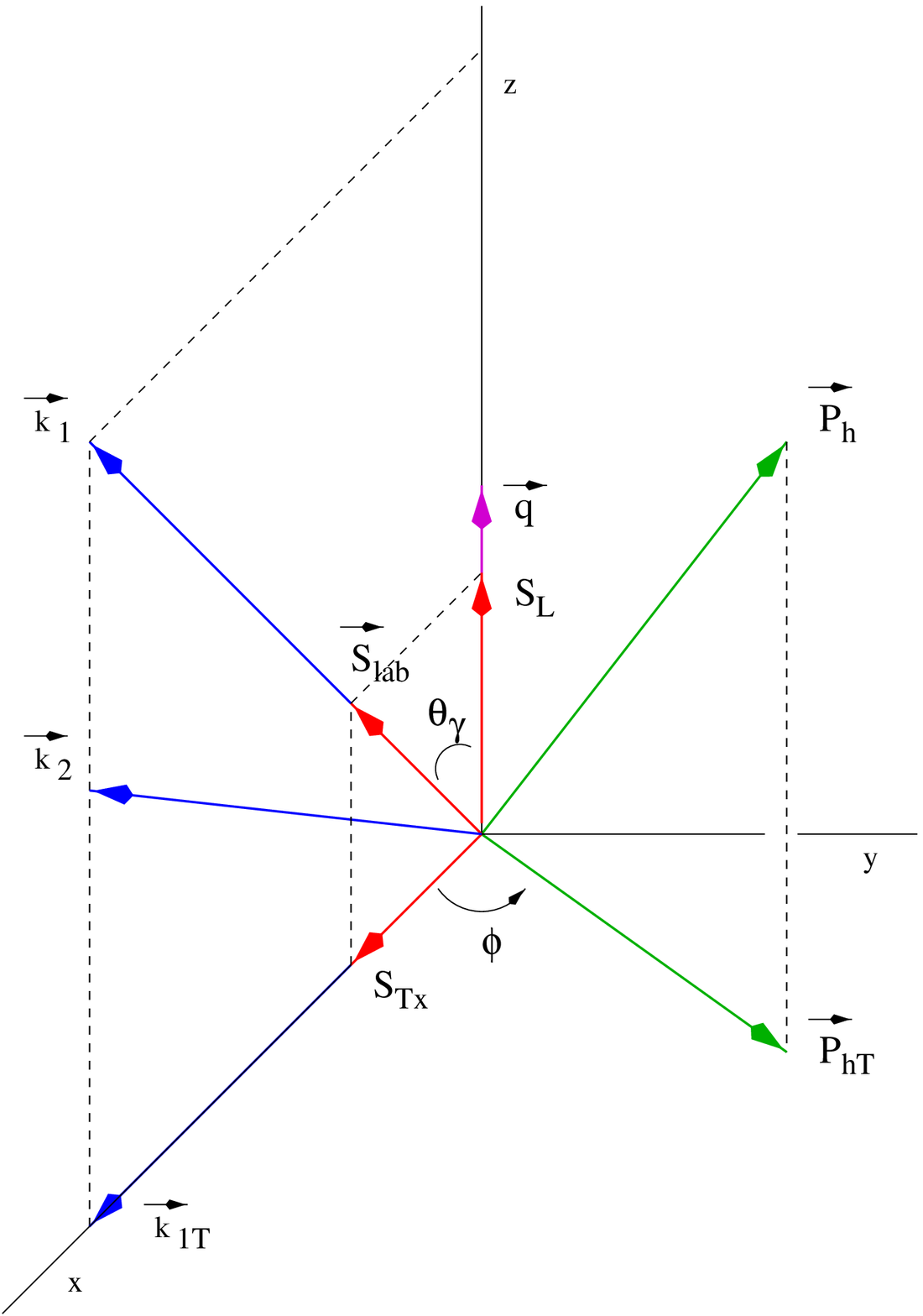,width=6.5cm,height=7.8cm}
{\small Figure 1: The kinematics of SIDIS. }
\medskip
\end{wrapfigure}

Due to the non-zero parton intrinsic transverse momentum, in the SIDIS 
cross section besides the conventional twist-2 non-perturbative blocks, 
there are different combinations of twist-two and twist-three structures, 
which could be probed in azimuthal asymmetries of hard scattering processes. 
The complete tree-level description expression containing contributions 
from twist-two and twist-three distribution and fragmentation functions in 
SIDIS has been given in Ref.~\cite{TM}. The full result for 
the SIDIS cross section contains a large number of terms. In this respect  
it is more practical to split up it in the parts involving the lepton 
polarizations, unpolarized (U) or longitudinally (L) polarized and target 
polarizations, unpolarized (U), longitudinal (L), and transverse (T) polarized 
keeping only the $\phi$-independent and $\phi$-dependent terms, relevant in  
following. It can be presented in the following 
way\footnote{Up to the $1/Q$ 
order.}: 
\begin{eqnarray}
d \sigma^{eN \to e h X} \propto  
& & d \sigma^{(0)}_{UU} + {1 \over Q} \cos \phi d \sigma^{(1)}_{UU}
\nonumber \\  
&+& {1 \over Q} \sin \phi \lambda d \sigma^{(2)}_{LU} 
\nonumber \\
&+& {1 \over Q} \sin \phi S_L d \sigma^{(3)}_{UL} 
+ \sin 2\phi S_L d \sigma^{(4)}_{UL} 
\nonumber \\
&+& \sin (\phi+\phi_S) S_T d \sigma^{(5)}_{UT} 
+ {1 \over Q} \sin (2\phi-\phi_S) S_T d \sigma^{(6)}_{UT} 
+ \sin (3\phi-\phi_S) S_T d \sigma^{(7)}_{UT}  \nonumber \\
&+& \lambda S_L d \sigma^{(8)}_{LL} + {1 \over Q} \lambda S_L \cos \phi 
d \sigma^{(9)}_{LL} \nonumber \\ 
&+& \lambda S_T \cos (\phi-\phi_S) d \sigma^{(10)}_{LT} 
+ {1 \over Q} \lambda S_T \cos (2\phi-\phi_S) d \sigma^{(11)}_{LT},  
\label{CS} 
\end{eqnarray}
where the first subscript corresponds to beam polarization and the second 
one to the target polarization. Here the terms proportional 
to $1/Q$ indicate the ``kinematical'' or dynamical twist-3 contributions. 

\begin{figure}
\epsfig{figure=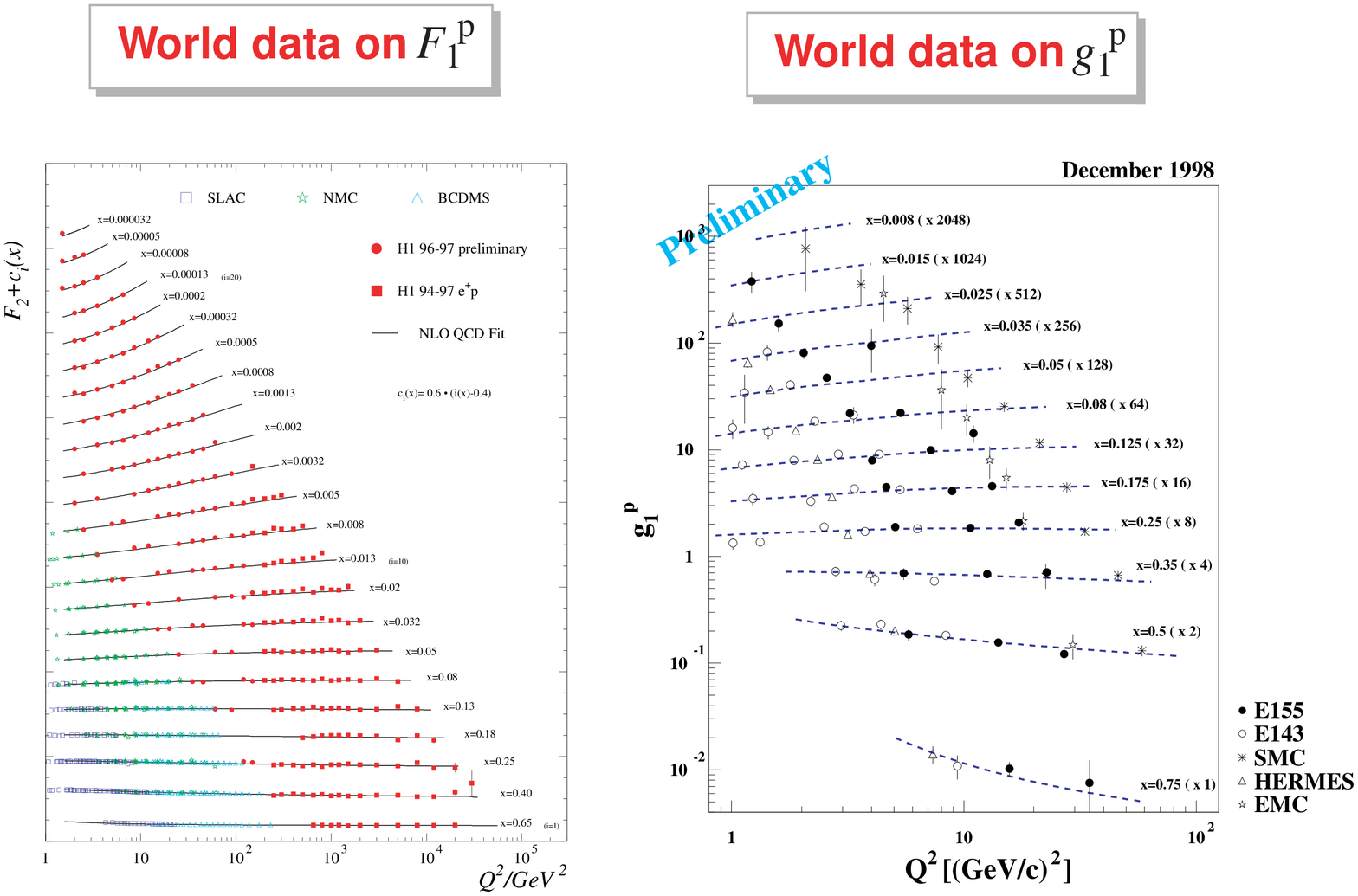,width=14.0cm,height=8.0cm}
{\small Figure 2: Word data on spin-average and helicity structure function 
from Ref.\cite{Makins}}
\medskip
\end{figure}

In inclusive processes, at leading $1/Q$ order, besides the well-known parton 
distribution $f_1(x)$, often denoted as $q(x)$, the longitudinal spin distribution 
$g_1(x)$, often denoted as $\Delta q(x)$, there is a third 
twist-two distribution function, the transversity distribution function $h_1(x)$, 
also often denoted as $\delta q(x)$. It was first discussed by Ralston and 
Soper\cite{RS} in double transverse polarized Drell-Yan scattering. The 
transversity distribution $h_1(x)$ measures the probability to find a transversely 
polarized quark in a transversely 
polarized nucleon. It is equally important for the description of the spin structure 
of nucleons as the more familiar function $g_1(x)$; their information being complementary. 
In the non-relativistic limit, where boosts and rotations commute, $h_1(x) = g_1(x)$; then  
difference between these two functions may turn out to be a measure for the relativistic 
effects within nucleons. On the other hand, there is no gluon analog on $h_1(x)$. 
This may have interesting consequences for ratios of transverse to longitudinal 
asymmetries in polarized hard scattering processes (see e.g. Ref.~\cite{JJJ}). 

The transversity remains still unmeasured, contrary to the 
case for spin-average and helicity structure functions, which are known over large 
ranges of $Q^2$ and $x$ (see Fig.2) The reason is 
that it is a chiral odd function, and consequently it is suppressed in inclusive 
deep inelastic scattering (DIS)~\cite{JJ,ArtruMek}. Since electroweak and strong interactions 
conserve chirality, $h_1(x)$ cannot occur alone, but has to be accompanied by a 
second chiral odd quantity. It is illustrated in Fig. 3 from Ref.~\cite{JAFFE}.

\begin{figure}
\epsfig{figure=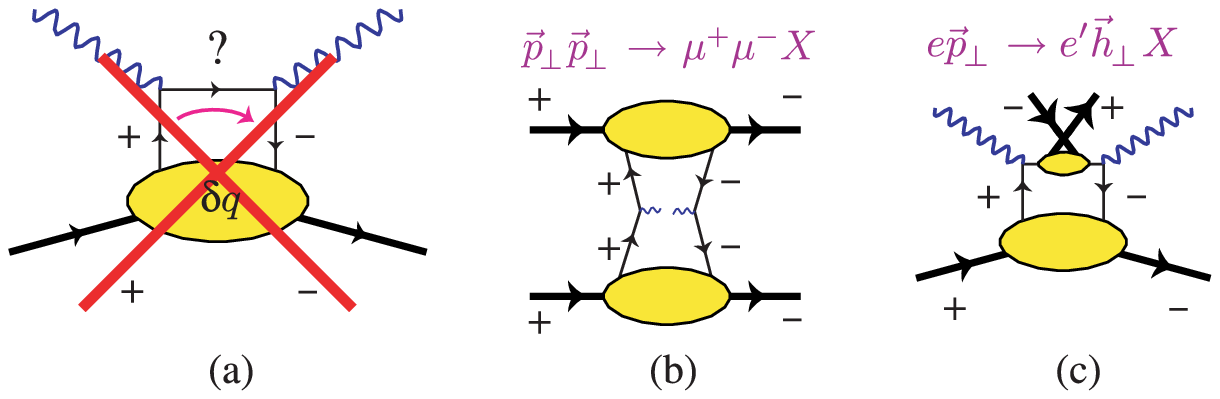,width=12.0cm,height=3.0cm}
{\small Figure 3: Deep inelastic processes relevant to transversity. }
\medskip
\end{figure}

In principle, transversity distributions can be extracted from cross 
section asymmetries in polarized processes involving a transversely 
polarized nucleon. In the case of hadron-hadron scattering these asymmetries 
can be expressed through a flavor sum involving a product of two chiral-odd 
transversity distributions. This is one of the main goals of the spin program 
at RHIC~\cite{RHIC} (Fig. 3b). 

In the case of SIDIS off transversely polarized nucleons (Fig.3c) there exist 
several methods to access 
transversity distributions. One of them, the twist-3 pion production 
\cite{jaffe-ji93}, uses longitudinally polarized leptons and measures a double spin 
asymmetry. Other methods do not require a polarized beam, and rely on the  
{\it polarimetry} of the scattered transversely polarized quark. They consist on: 
\begin{itemize}
 \item the measurement of the transverse polarization of $\Lambda$'s in the 
       current fragmentation region \cite{jaffe96,ArtruMek},
 \item the observation of a correlation between the transverse spin vector
       of the target nucleon and the normal to the two-meson plane 
       \cite{jaffe97b,jaffe97a},
 \item the observation of the ``Collins effect'' in quark fragmentation through 
       the measurement of pion single target-spin asymmetries
       \cite{COL,AK,TM}.  
\end{itemize}

In Sec. 2 I will focus on the last method -- Collins fragmentation 
function, $H_1^{\perp}$ which can be simply interpreted as the production 
probability of an unpolarized hadron from a transversely polarized 
quark~\cite{COL}. A first indication of a nonzero $H_1^{\perp}$ comes from 
analysis of the 91-95 LEP1 
data (DELPHI)~\cite{EST}~\footnote{Possibilities of measuring $H_1^{\perp}$ at 
BELLE~\cite{BELLE} are currently being examined.}. In numerical calculations 
the Collins ansatz~\cite{COL} for the analyzing 
power of transversely polarized quark fragmentation was used:
\begin{equation}
A_C(z,k_T) \equiv \frac{\vert k_T \vert}{M_h}\frac{H_1^{\perp}(z,k_T^2)}
{D_1(z,k_T^2)} = \eta \frac{M_C\,\vert k_T \vert}{M_C^2+k_T^2},
\label{H1T}
\end{equation}
where $\eta$ is taken as a constant, although, in principle it could be $z$ 
dependent. In this case the $z$ dependence of the single-spin azimuthal  
asymmetries (SSAA) conditioned by strong correlations between kinematical 
variables.     

Due to non-zero quark transverse momentum in semi-inclusive processes, at leading 
$1/Q$ order, in addition to the above discussed distribution functions, there are 
three non-vanishing distribution functions, $g_{1T}, h^{\perp}_{1L}, h^{\perp}_{1T}$. 
These functions relate the transverse (longitudinally) polarization of the quark to 
the longitudinally (transverse) polarization of the nucleon. 

\section{Single-spin azimuthal asymmetries}

In recent years significant SSAA have been observed in experiments with transversely 
polarized proton and anti-proton beams, respectively\cite{E704}.

Very recently a significant target-spin asymmetry of the distributions in the azimuthal 
angle $\phi$ of the pion related to the lepton scattering plane for $\pi^{+}$ 
electroproduction in a {\it longitudinally} polarized hydrogen target has been 
observed by the HERMES collaboration~\cite{HERM,HERM1} (Fig.4). At the same 
time the SMC collaboration has studied the azimuthal distributions of pions 
produced in deep inelastic scattering off {\it transversely} polarized protons 
and deuterons~\cite{SMC}.   

\begin{wrapfigure}{l}{9cm}
\epsfig{figure=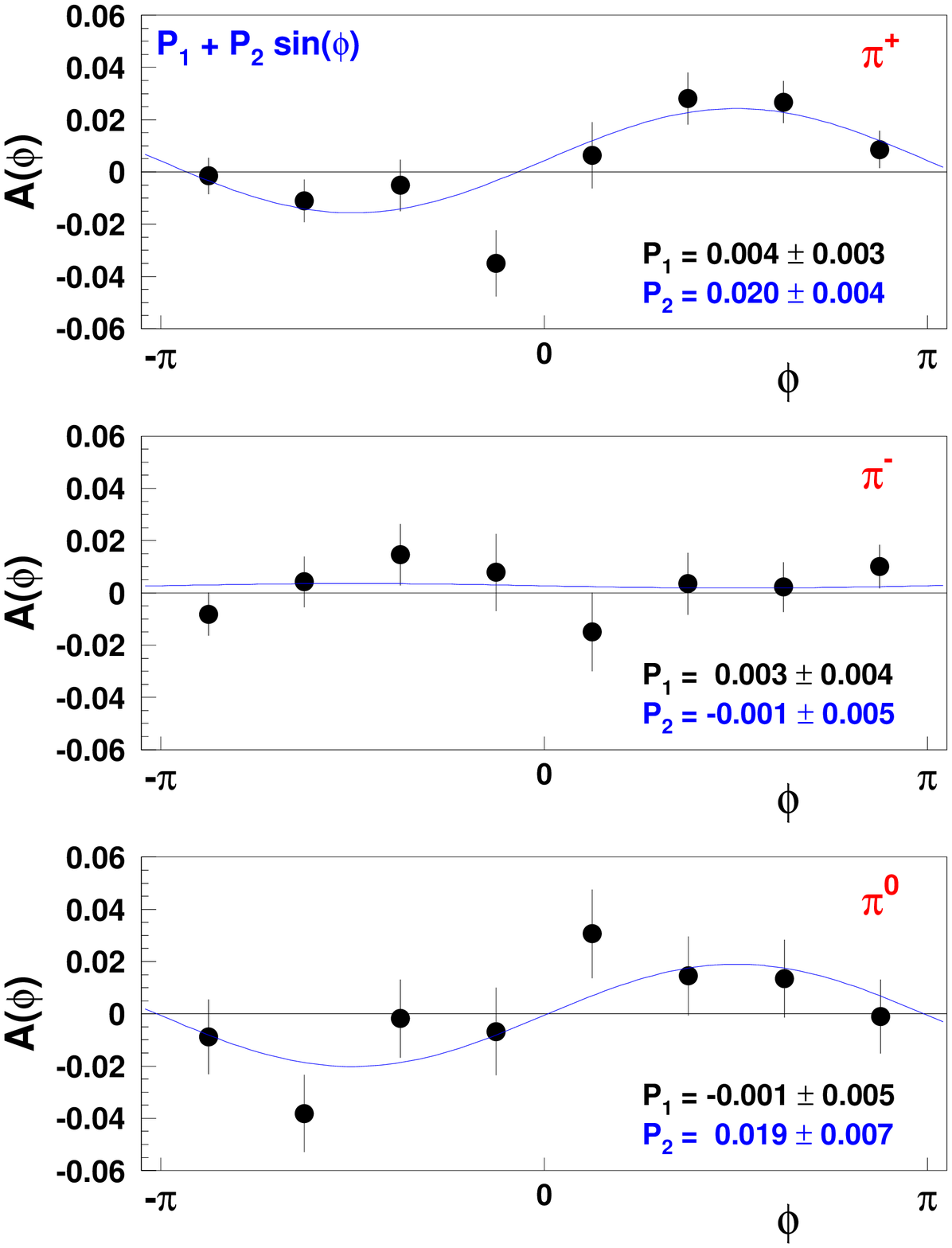,width=7.5cm,height=7.5cm}
{\small Figure 4: 
$ A(\phi) = {1 \over \langle P \rangle } \frac{N^{+}(\phi) - 
N^{-}(\phi)}{N^{+}(\phi) + N^{-}(\phi)}$, $N^{\pm}$ is a number of events with 
positive/negative target helicity; $\langle P \rangle$ -- average target 
polarization. Data are from Refs.~\cite{HERM,HERM1}.
}
\medskip
\end{wrapfigure}

These non-zero asymmetries may originate from multi-parton correlations in 
initial or final states and non-zero parton transverse momentum.  
They have initiated a number of phenomenological approaches to evaluate 
these asymmetries using different input distribution and fragmentation 
functions. An analysis of different approximations, which aim at 
explaining the experimental data, have been provided in 
Ref.~\cite{OBDN}. 
The approximation where the twist-2 {\it transverse} quark spin 
distribution in the {\it longitudinally} polarized nucleon, 
$h_{1L}^{\perp(1)}(x)$, is considered small enough to be 
neglected~\cite{BBE,DNO} with the assumption of the u-quark dominance are in good 
agreement with the Bjorken-$x$, $z$, and $P_{hT}$ behaviors of the 
{\it sin}$\phi$ asymmetry for charged and neutral pion production observed 
at HERMES (see Figs. 5,6)~\footnote{For $\pi^{-}$ production the data 
are consistent with zero in agreement with the result of the approximation.}.  
Note, that it does not require the twist-3 interaction-dependent part 
of the fragmentation function, $\tilde{H}(z)$, to be zero, which leads 
to the inconsistency that all T-odd fragmentation functions would be 
required to vanish~\cite{TM,ST}. 

Results on SSAA provide evidence in  support of the existence of non-zero 
chiral-odd structures that describe the transverse polarization of quarks. 
New data are expected from future HERMES, COMPASS measurements on a
transversely polarized target, which will give direct access to the 
transversity~\cite{KNO}. 

\begin{figure}
\epsfig{figure=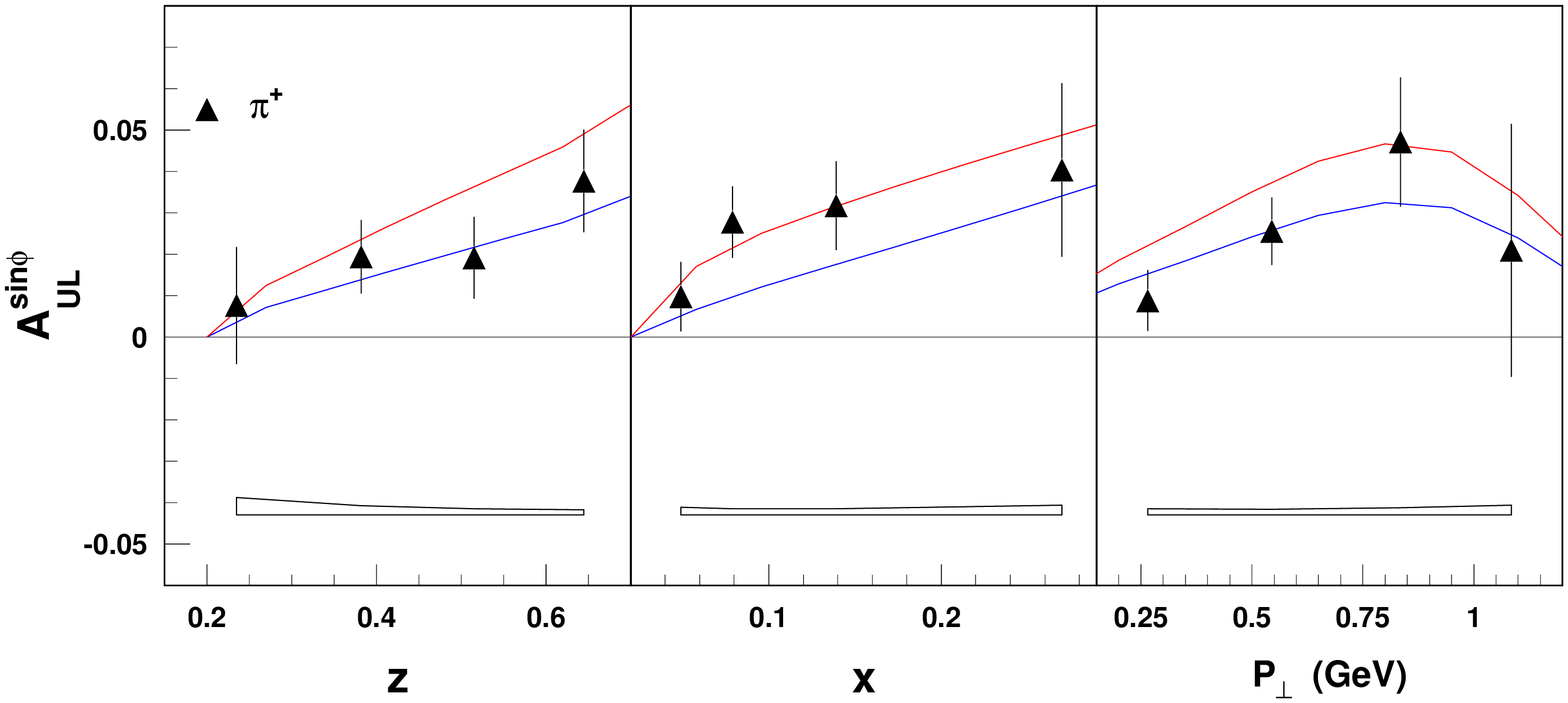,width=14.0cm,height=8.0cm}
{\small Figure 5: $A_{UL}^{\sin \phi}$ for $\pi ^+$ electroproduction as 
a function of $z$, the Bjorken-$x$, and of the $P_{hT}$ from Ref.~\cite{HERM}. 
Error bars include the statistical uncertainties only. The open bands at 
the bottom of the panel represent the  systematic uncertainty. 
The curves show the range of predictions of a model calculation~\protect\cite{DNO}.
}
\medskip
\end{figure}

\begin{figure}
\epsfig{figure=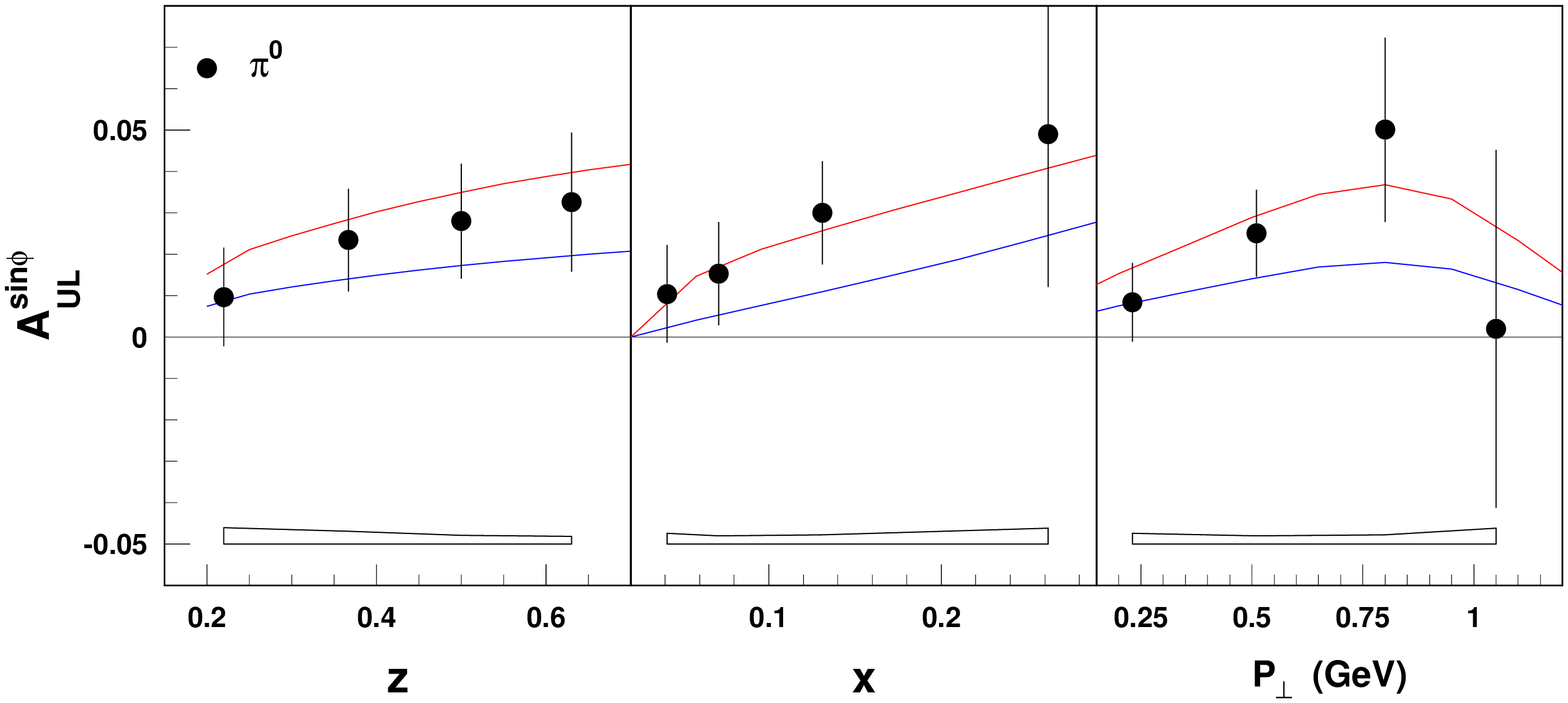,width=14.0cm,height=8.0cm}
{\small Figure 6: The same as in Fig.6 for neutral pion electroproduction. 
Data are from Ref.~\cite{HERM1}. The curves show the range of predictions 
of a model calculation~\protect\cite{DNO}.}
\medskip
\end{figure}

\section{Spin-independent azimuthal asymmetries}

The nonzero transverse momenta of the partons as a consequence 
of being confined by the strong interactions inside hadrons, generates 
significant azimuthal asymmetries, which show up as a {\it cos}$\phi$ and 
{\it cos}$2\phi$ terms in the {\it unpolarized} SIDIS~\cite{CAHN}. 
The experimental results of the EMC~\cite{EMC} and E665~\cite{E665} 
are described well by the simple parton model~\cite{KKCK}, 
where both first-order perturbative QCD effects~\cite{GP} and 
non-perturbative intrinsic transverse momentum effects~\cite{CAHN} were 
taken into account. The most important contribution to azimuthal 
dependence of {\it unpolarized} SIDIS comes from kinematical effect of 
the intrinsic transverse momentum. However, to get a  
complete behavior of azimuthal distributions one has to take into account 
also higher twist effects~\cite{BERG,TM}. In the Ref.~\cite{BKM} authors tried  
to isolate the higher twist effects. A self-consistent study of 
unpolarized azimuthal asymmetries. i.g. taking into account all effects 
that generate these asymmetries is still an open issue. 
To make an analogy between spin-independent and double-spin {\it cos}$\phi$ 
asymmetries I would like to present the tree level contributions to 
spin-independent {\it cos}$\phi$ asymmetry from Ref.~\cite{TM}:    
$$
d \sigma \propto \quad   
d \sigma^{(0)}_{UU}
+ {1 \over Q} \cos \phi d \sigma^{(1)}_{UU}  
$$
being 
\begin{equation}
d \sigma^{(0)}_{UU} 
\propto  \sum_a e^2_a f_1^a(x) D_1^a(z),
\label{fd}  
\end{equation}
\begin{equation}
d \sigma^{(1)}_{UU} 
\propto  - \sum_a e^2_a \Biggl ( f^{\perp (1) a}(x) D_1^a(z) - f_1^a(x) 
\tilde{D}^{\perp (1) a} \Biggr )  , 
\label{UUC}
\end{equation}
where 
$$
f^{\perp (1)}(x)  = f_1^{(1)}(x)/x + \tilde{f}^{\perp (1)},
$$
$$
\tilde{D}^{\perp (1)}(z) = D^{\perp (1)}(z) - z D_1^{(1)}(z). 
$$

When in the Eq.(\ref{UUC}) the interaction dependent parts of DF's and 
FF's are set to zero, the asymmetry reduces to a kinematical effect 
conditioned by intrinsic transverse momentum of partons in the nucleon 
as was calculated by Cahn~\cite{CAHN}. 

\section{Double-spin azimuthal asymmetries} 

Here I will focus my attention on the {\it cos}$\phi$ moment of the double-spin 
azimuthal asymmetry (DSAA) 
for pion electroproduction in semi-inclusive deep inelastic scattering of 
longitudinally polarized leptons off longitudinally polarized protons 
(for details see Ref.~\cite{DMO}). The contribution to double-spin {\it cos}$\,\phi$ 
asymmetry with the combinations of different leading and sub-leading distribution 
and fragmentation function can be symbolically presented in the following way  
\begin{eqnarray}
d \sigma \propto  
& & \lambda S_L d \sigma^{(8)}_{LL} \nonumber \\ 
&+& {1 \over Q} \lambda S_L \cos \phi 
d \sigma^{(9)}_{LL} 
\nonumber \\ 
&+& \lambda S_T \cos \phi 
\sin \theta_{\gamma} 
d \sigma^{(10)}_{LT}, \label{DS} 
\end{eqnarray}
where
\begin{equation}
d \sigma^{(8)}_{LL} \propto  \sum_a e^2_a 
\,g^a_1(x)\, D^a_1(z),  
\end{equation}
\begin{equation}
d \sigma^{(9)}_{LL} \propto \sum_a e^2_a     
\Biggl (g^a_1(x)\,\tilde{D}^{\perp a (1)}(z) - 
g_L^{\perp a (1)}(x) D^a_1(z)  
- h_{1L}^{\perp (1) a}(x) \tilde{E}^{a}(z)  \Biggr ),  
\end{equation}
\begin{equation}
d \sigma^{(10)}_{LT} \propto 
\sum_a e^2_a g_{1T}^{a (1)}(x)\,D^{a}_1(z).   
\end{equation}

The {\it cos}$\,\phi$ DSAA in the SIDIS 
cross-section can be defined as appropriately weighted integral 
of the cross section asymmetry:
\begin{equation}
< {\vert P_{hT}\vert} \cos \phi>_{LL} = 
\frac{\int d^2P_{hT} {\vert P_{hT}\vert}
\cos \phi \left(d\sigma^{++}+d\sigma^{--}-d\sigma^{+-}-d\sigma^{-+} \right)}
{{1 \over 4}\int d^2P_{hT} \left(d\sigma^{++}+d\sigma^{--}+
d\sigma^{+-}+d\sigma^{-+} \right)}. 
\label{ASMY}
\end{equation}
Here $++,-- (+-,-+)$ denote the anti-parallel (parallel) polarization of the 
beam and target~\footnote{this leads to positive $g_1(x)$.} and $M_h$ is the mass of 
the final hadron. The above defined weighted asymmetry is related to the 
experimentally observable asymmetry through the following relation 
\begin{equation}
A_{LL}^{\cos\phi} \approx {1 \over {\langle P_{hT} \rangle}} 
 {\langle {{\vert P_{hT}\vert} \cos \phi \rangle}}_{LL}.  
\label{AS0}
\end{equation}
Using the Eqs.( \ref{CS}), (\ref{DS}) one can get 
\begin{equation}
A_{LL}^{\cos\phi} = 4 \frac{d \sigma^{(9)}_{LL} + 
\sin{\theta}_{\gamma} d \sigma^{(10)}_{LT}}{d \sigma^{(0)}_{UU}}.  
\label{AS}
\end{equation}

\begin{wrapfigure}{l}{9cm}
\epsfig{figure= 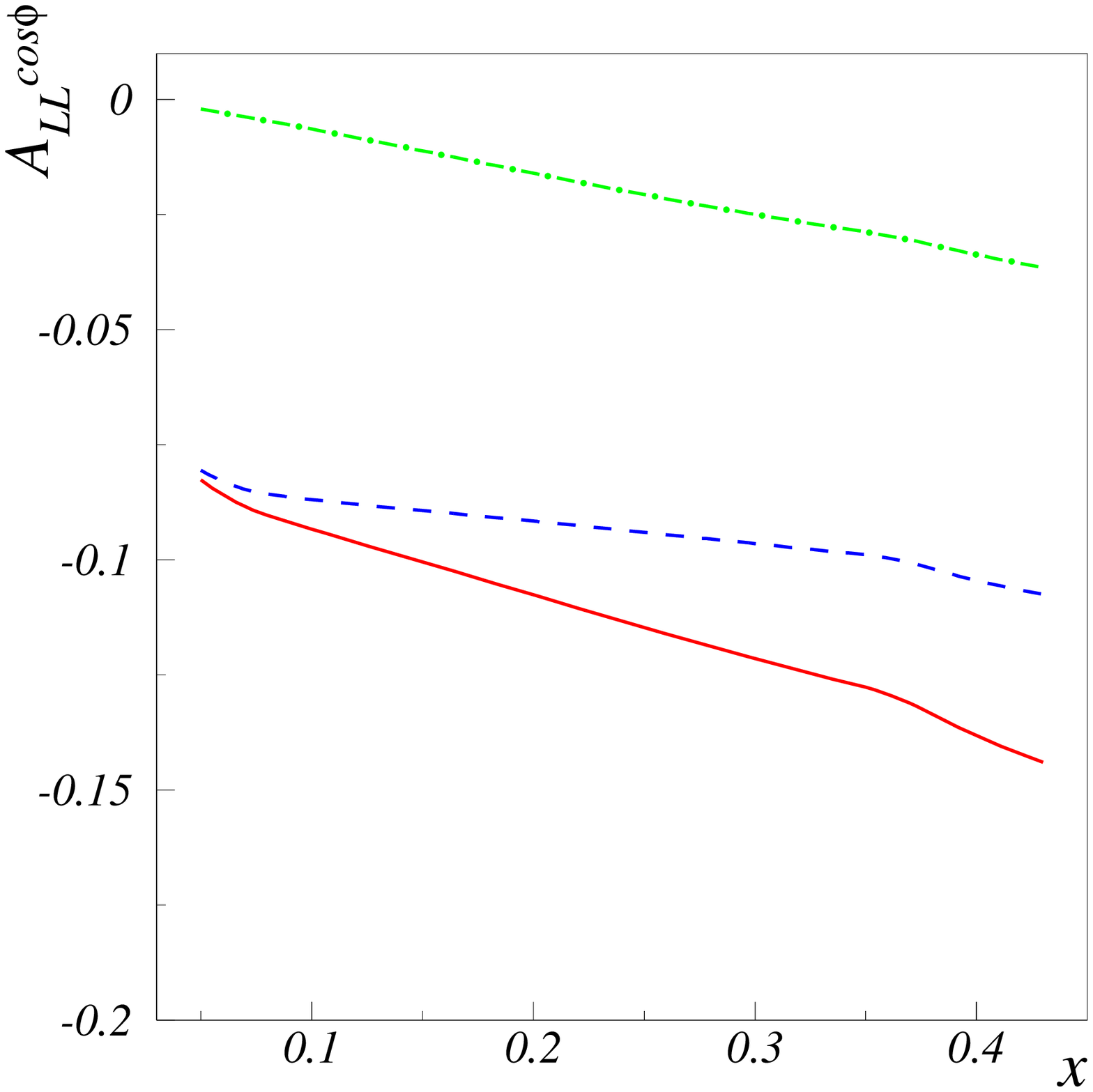,width=8.0cm,height=8.0cm}
{\small Figure 7: $A^{\cos\phi}_{LL}$ for $\pi^{+}$ 
production as a function of Bjorken $x$. The dashed line corresponds to 
contribution of the $d \sigma^{(9)}_{LL}$, dot-dashed one to 
$\sin{\theta}_{\gamma} d \sigma^{(10)}_{LT}$ and 
the solid line is their sum.}
\medskip
\end{wrapfigure}

To estimate that asymmetry we take into account only the $1/Q$ order 
contribution to the DSAA which arises from intrinsic $p_T$ 
effects~\cite{AK} similar to the $\cos\phi$ asymmetry in {\it unpolarized} 
SIDIS~\cite{CAHN}, i.e.  all twist-3 interaction dependent distribution and 
fragmentation functions are set to zero~\cite{DMO}. In that approximation 
assuming a Gaussian parameterization for the distribution of the initial 
parton's intrinsic transverse momentum, $p_T$, in the helicity distribution 
function $g_1(z,p^2_T)$ one obtains  
\begin{equation}
g_1^{(1)}(x) = \frac{\pi <p_T^2>}{2M^2} g_1(x). 
\label{g1LL}
\end{equation}

To estimate the transverse asymmetry contribution ($d \sigma_{LT}^{(10)}$) into 
the $A_{LL}^{\cos \phi}$, one can act as in the Ref.~\cite{KM}. 

In Fig.7, the asymmetry $A^{\cos\phi}_{LL}(x)$ of Eq.(\ref{AS}) 
for $\pi^{+}$ production on a proton target is presented as a function of
$x$-Bjorken. The curves are calculated by integrating over the HERMES kinematic 
ranges, taking $\langle P_{hT} \rangle = 0.365$ GeV as input (for more details 
see Ref.~\cite{DMO}).  

From Fig.7 one can see that the approximation where all twist-3 DF's 
and FF's are set to zero gives the large negative double-spin {\it cos}$\phi$ 
asymmetry at HERMES energies. Note that the `kinematic` contribution to 
$A^{\cos\phi}_{LL}(x)$ coming from the transverse component of the target 
polarization, is negligible. 

\section{Conclusions}

The leading and sub-leading order results for pion electroproduction in 
polarized and unpolarized semi-inclusive 
deep-inelastic scattering, putting emphasis on transverse 
momentum dependent effects appearing in azimuthal asymmetries is considered. 
In particular, the spin-dependent (single, double) and spin-independent 
asymmetries of the distributions in the azimuthal angle $\phi$ 
of the pion related to the lepton scattering plane are discussed. 

It is shown that the approximation where the twist-2 {\it transverse} 
quark spin distribution in the {\it longitudinally} polarized nucleon, 
$h_{1L}^{\perp(1)}(x) \approx 0$ with the assumption of the u-quark 
dominance, gives a consistent description of recent HERMES data on SSAA.  

In {\it spin-independent} SIDIS the different mechanisms 
which generate {\it cos}$\phi$ asymmetry is discussed. 
At moderate $Q^2$ and small $P_{hT}$ the main contribution 
to the asymmetry comes from kinematical $p_T$ effects 
describing well the existing experimental results.  

At HERMES kinematics a sizable negative {\it cos}$\phi$  
double-spin asymmetry for $\pi^{+}$ electroproduction in SIDIS 
is predicted taking into account only the $1/Q$ order contribution 
to the DSAA which arises from intrinsic $p_T$ effects similar to the 
{\it cos}$\phi$ asymmetry in unpolarized SIDIS: 
all twist-3 interaction dependent distribution and fragmentation functions  
are set to zero. The ``kinematical'' contribution from target 
transverse component ($S_{Tx}$) is well defined and it is shown that its 
contribution to $A_{LL}^{\cos \phi}$ is negligible. 
The double-spin {\it cos}$\phi$ asymmetry is a good observable to investigate 
the weights of twist-2 and twist-3 contributions at moderate $Q^2$. 
It may give a possibility to estimate $<p^2_T>$ of partons in the nucleon. 
The complete behavior of azimuthal distributions may be predicted only after 
inclusion of higher-twist and pQCD contributions, nevertheless, if one consider 
the kinematics with $P_{hT} < 1$ GeV and $z < 0.8$, one can isolate the 
non-perturbative effects conditioned by the intrinsic transverse momentum 
of partons in the nucleon.  

\section{Acknowledgments} 

I would like to thank M.~Anselmino, D.~Boer, E.~De~Sanctis, 
A.~Efremov, R.~L.~Jaffe, R.~Jakob, A.~Kotzinian, P.~J.~Mulders, 
W.-D.~Nowak, A.~Sch\"afer, and O.~Teryaev for many useful and 
stimulating discussions.

\end{document}